Torgeir Dingsøyr and Yvan Petit
# 4 Managing layers of risk: Uncertainty in large development programs combining agile software development and traditional project management

## 4.1 Introduction

Software development projects have a track record of schedule and cost overruns, and have often faced challenges with delivering expected quality (Flyvbjerg & Budzier, 2011). In the nineties many expressed concern about software projects and used the word "crisis" (Kraut & Streeter, 1995). Consequently, project management professional associations and some authors have proposed that software development projects should adopt and implement the traditional risk management approaches as a contributor to success (Dey et al., 2007). This is quite natural since risk management has been considered one of the core knowledge areas in project management for many decades. Literature abounds in this field (Chapman & Ward, 2003; Jaafari, 2001; Johansen et al., 2019; Kendrick, 2015; Persson et al., 2009; Project Management Institute, 2019; Raz et al., 2002) and most general project management books include at least a section on risk management (Andersen, 2008; Dinsmore & Cabanis-Brewin, 2014; Gray & Larson, 2014; Kerzner, 2017; Nicholas, 2017). Risk management has also been shown to bring a number of benefits such as identification of favourable alternative courses of action, reduced surprises, and provided more precise estimates (Bannerman, 2008). However, recent research has also shown that these practices are not widely used in software development projects (Bannerman, 2008; Odzaly & Des Greer, 2014).

Risk Management is covered in the PMBOK Guide® (Project Management Institute, 2017) which defines a project risk as: "an uncertain event or condition that, if it occurs, has a positive or negative effect on a project objective" (p. 720) with processes for risk identification, risk categorization, risk qualitative and quantitative analysis, plan risk responses, monitor risks and implement risk response. Both the Project Management Institute (PMI) and Association for Project Management (APM) define a risk as an uncertain "event" which might have positive effects (opportunities) or negative effects (threats) (Association for Project Management, 2008). In general, however, project managers and the literature tend to focus on threats rather than on opportunities (Johansen, 2015). One of the most used textbooks in software engineering, for example, defines risks as "something you´d prefer not to happen" (Sommerville, 2016). Several techniques have been developed to assess the probability of occurrence and the potential impacts to projects. A typical classification of risks is based on the level of knowledge about the possibility for the risk to take place (known or unknown) and the level of knowledge about the impact (known or unknown).

In projects undertaken in rapidly changing environments where uncertainty may be unavoidable, such as software development projects, managers need to go beyond traditional risk management; adopting roles and techniques oriented less towards planning and more towards flexibility and learning (De Meyer et al., 2002; Loch et al., 2006; Pich et al., 2002; Platje & Seidel, 1993). Some authors have therefore advocated for the use of the broader concept of *uncertainty management* instead of *risk management* (Cleden, 2009; Ibrahim et al., 2009; Perminova et al., 2007, 2008; Ward & Chapman, 2003). "Uncertainty management is not just about managing perceived threats, opportunities and their implications. [. . .] It implies exploring and understanding the origins of project uncertainty before seeking to manage it, with no preconceptions about what is desirable or undesirable" (Ward & Chapman, 2003, p. 98–99). An uncertainty management perspective draws attention to the need to understand variability in future activities; the individual's inability to assign probabilities to events, and their inability to predict accurately what the outcomes of a decision might be (Duncan, 1972). This has been shown to be particularly important in safety-critical projects (Saunders, 2015a, 2015b; Saunders et al., 2016). Dönmez and Grote (2018) summarize the difference between risk and uncertainty as follows: "risk refers to an unknown event that leads to one outcome from a set of known outcomes, each of which can be assigned a probability (however estimated). Uncertainty, in contrast, relates to a lack of knowledge about which outcomes are possible, including both their nature and associated probabilities. An 'uncertainty' is thus an unknown event from an unknown set of possible outcomes" (p. 95).

Uncertainty might apply to multiple facets of a project and some authors have tried to define categories of uncertainties. This ranges from simple models with two categories such as endogenous vs. exogenous uncertainty (Ahsan et al., 2010), to more extensive models. A review of uncertainty in project management by Jalonen (2011) uses the following categories: technology, markets, regulation, social/ political, acceptance/legitimacy, managerial, timing, and consequence uncertainty. More suitable to the software development projects are the three categories requirement uncertainty, resource uncertainty and task uncertainty proposed by Dönmez and Grote (2018) or the model by Ropponen and Lyytinen (2000) which was used in this study as it is widely used in previous studies to categorize software project risks: scheduling and timing, system functionality, subcontracting, requirement management, resource usage and performance and personnel management. Such categorization helps practitioners and researchers to identify adequate responses and techniques according to the uncertainty category (Cagliano, 2015).

Through multiple short iterations in conjunction with more frequent and earlier feedback, agile approaches such as Scrum (Hossain et al., 2009; Schwaber, 2004; Schwaber & Sutherland, 2013) have attempted to implicitly address the first category of uncertainty i.e., requirement uncertainty (Tuunanen et al., 2015). However, a qualitative study from 2014 (Siddique & Hussein, 2014) reveals that many Agile practitioners handle risk as in a traditional waterfall approach. Some authors also claim that, while having addressed the concern with requirement uncertainty, additional and new risks have also been introduced by the adoption of agile approaches such as new development cycle risks, development environment risks and programmatic risks (Walczak & Kuchta, 2013). Although agile approaches have brought product owners (or clients) closer to development there remains a separation of development and operations. Agile has also exacerbated the impact of technical debt[1] (Kruchten et al., 2012).

Agile approaches have initially been developed for small development projects with one team but recently they have been scaled to include multiple teams in larger organizations (Leffingwell, 2015; Vaidya, 2014), for example in the oil industry (Grewal & Maurer, 2007), in large software development organizations (Bick et al., 2018; Gruver & Mouser, 2015; Lindvall et al., 2004), in regulated environments (Fitzgerald et al., 2013) and in the public sector (Dingsøyr et al., 2018a). Software development projects might vary from simple one-team development to very large-scale projects with more than ten teams (Dingsøyr et al., 2014). However, studies on project success suggest that agile methods outperform traditional methods in contexts with high dynamism (Butler et al., 2019) and both in small and large projects (Jørgensen, 2019). Scaling agility have introduced a range of new challenges which have only been recently studied (Conboy & Carroll, 2019; Hobbs & Petit, 2017a, 2017b), including approaches to uncertainty management.

In general, we have few recommendations on risk management for large software projects. Sommerville´s textbook on software engineering (Sommerville, 2016) recommends establishing a risk register with consequence analysis and to establish a risk management plan which should be revisited throughout the project. A practice which is recommended for risk analysis is to identify and monitor the "top 10 risks". However, the author notes that in agile development, risk management is "less formal". Agile methods are believed to reduce risks related to requirements, but, on the other hand, increase risks related to staff turnover as documentation is more limited and communication more informal.

"Agile software development teams rely on high levels of autonomy. This has implications for the way in which they approach uncertainty. When left without clear guiding structures, their choice regarding how to deal with uncertainty becomes an uncertainty itself" (Dönmez & Grote, 2018). As a consequence, in contexts of multi-team projects, some additional levels of monitoring and coordination must therefore be introduced. This applies to all types of activities, including risk/uncertainty management. Inter-team coordination could be done by standards (using the rules by which something is done), by plans (achieved by specifying what is to be produced, by whom and when), by formal mutual adjustment, and by informal mutual adjustment (Dingsøyr et al., 2018b; McBride, 2008; Sabherwal, 2003).

---

[1] Technical debt has been introduced as a metaphor in software development, describing the cost of taking an easy solution now that is cheaper to implement than a solution which will be more robust over time.

Many of the strategies to manage risks in Agile development are implicit (Hijazi et al., 2012; Nyfjord & Kajko-Mattsson, 2007; Odzaly & Des Greer, 2014). Moran argues that "risk management in agile projects remains a passive and implicit activity that can be misdirected and often misunderstood [. . .] whilst most developers have little difficulty explaining which features they are working on (e.g., user stories) or to what level of quality they should be completed (e.g. definitions of done), few can comment on the capacity of their work to reduce (or exploit) project risk" (Moran, 2014, p. 33). There is some disagreement as to whether explicit methods of risk management should be used in projects which are executed according to Agile methodology or if the implicit risk management built into Agile methodologies is sufficient (Walczak & Kuchta, 2013), some authors argue that this might depend on the context (Howell et al., 2010; MacCormack & Verganti, 2003).

Most software development projects are today using agile methods, including large projects with multiple development teams. The special issue on large-scale agile development in *IEEE Software* (Dingsøyr et al., 2019) describes a transition from first generation large-scale agile methods such as advice from project management frameworks like Prince2 combined with Scrum, to second generation frameworks such as the Disciplined Agile Delivery, Scaled Agile Framework and Large-Scale Scrum. These new frameworks add practices, roles and new artefacts to manage software development. An example of a program using a first generation large-scale agile development method, the "Perform" programme for the Norwegian State Pension Fund was organised in four main projects focusing on establishing business requirements, the technical architecture, development and test (Dingsøyr et al., 2018a). The rich description does not describe explicit practices for risk management, but the programme was organised into 12 releases, all development teams used three-week iterations and there were a number of arenas to ensure customer engagement and internal programme coordination.

Prior studies on uncertainty management in agile software development have focused on several types of uncertainty as discussed above (Dönmez & Grote, 2015). But we also find studies which focus on specific areas such as knowledge sharing, uncertainty in assessing value and cost and management of technical risks: To manage knowledge sharing risks in agile development, Ghobadi and Mathiassen (2017) have developed a model with seven risk areas such as "team diversity" and "project technology" and five areas of resolution strategies such as "strengthen resources" and "improve processes". This model is based on studies of four single-team projects. A practically oriented method to assess value and cost in agile software development projects is described in a magazine article by Hannay et al.

(2019). The method is described as suitable for large projects but does not focus on organisational layers. We also find studies which suggest models on how to handle technical risks as expressed in the software architecture by Nord et al. (2014).

The second-generation frameworks for agile development introduce a number of new practices, roles and artefacts to manage software development. With several teams working on the same project, new risks are introduced and advice on how to handle risks in single-team development may no longer be sufficient. We have tried to search literature for research on risk management at multiple layers i.e., how risks are managed at project level, release level, team level. Despite the fact that risk management has been researched for many decades, this specific topic seems to have been neglected, especially in projects using agile approaches. Nkukwana and Terblanche observed that "implementation teams value the governance role that project managers fulfil on agile projects, particularly with regards to project delivery, risk management, reporting and budgeting" (Nkukwana & Terblanche, 2017, p.6) but they have not tried to assess how this is done in practice.

How risks are managed implicitly and explicitly at multiple levels of agile projects has not been extensively studied (Nelson, 2008) and there is a need to investigate how risk management can be used in large agile projects (Odzaly & Des Greer, 2014). This is the objective of this exploratory study which investigates the following research question: *How does a large software/hardware development project using agile practices manage uncertainty at project/subproject and work package levels?* We believe this study of a first-generation agile development project will be important for researchers addressing uncertainty management in second-generation frameworks.

## 4.2 Method

We conducted an exploratory case study (Runeson & Höst, 2009; Yin, 2009) of a "very large" development project (Dingsøyr et al., 2014) of a new product, the Joint Strike Missile project at Kongsberg Defence & Aerospace. The case was selected as a project which combined traditional project management with agile development methods. The company had a traditional internal risk and opportunity process, and should be considered a "typical" project with respect to managing uncertainty. The case involved both hardware and software development. We chose a single-case study design in order to get a thorough understanding, as we could not find rich case descriptions on this topic in the literature and we were uncertain about the volume and range of practices employed. Single-case studies allow for generalisation of findings (Flyvbjerg, 2006) and can provide significant contributions to scientific development.

A survey distributed by the company amongst participants in the project indicated that uncertainty management was an area with potential for improved practices, as the score on the question "the project has plans for uncertainty management" was particularly low compared to other questions on project management. The survey was completed by 60 project participants and was followed up with a half-day workshop on uncertainty management where researchers provided recommendations from software engineering and project management literature. Data collection was done when the project was near completion of the third of four phases. We were granted access to interview 11 participants (see Table 4.1), three subproject managers, the assistant project manager (with responsibility for uncertainty management) – we refer to these in the following as managers. Further, four work package leaders from three subprojects (referred to as work package leaders) and finally three employees working in work packages in three subprojects (work package team members). This choice of informants, though a small number, allowed us to get diverse opinions on practices at the two levels.

**Table 4.1:** Overview of informants for study.

| Level | Assistant project manager | Subproject manager | Work package leader | Work package team member |
|---|---|---|---|---|
| Project | 1 | | | |
| Subproject1 | | 1 | 1 | 1 |
| Subproject2 | | 1 | 2 | 1 |
| Subproject3 | | 1 | 1 | 1 |
| **Total** | 1 | 3 | 4 | 3 |

The semi-structured interviews were conducted by the first author. We used the interview guide in Appendix 1 to get informants to describe their background, approaches to uncertainty management, and practices used in agile development. We also focused specifically on how they had managed opportunities so far in the project, as prior research has indicated a focus on threats. The interviews lasted from 30 to 45 minutes, a total of 120 pages of text were transcribed and sent back to informants for validation. We also included a quantitative second part, where we asked interviewees to indicate current and future importance of risk factors in the Ropponen and Lyytinen (2000) model.

Interview material was imported into NVivo for qualitative analysis, where the material was coded into groups that described explicit and implicit practices for uncertainty management at different levels of the project (see Table 4.2 for resulting categories). The project was located at two development sites, most subprojects were conducted at the main site, but some subprojects have work packages conducted at the other site which meant that possibilities for informal coordination was limited. Subprojects 1 and 2 were at the main site, while the work package leader and team members in Subproject 3 were from the other site.

Table 4.2: Mitigating practices to manage uncertainty at project and work package levels; divided into explicit practices and implicit practices.

| | Explicit | | Implicit | |
| --- | --- | --- | --- | --- |
| | Common | Specific | Common | Specific |
| Project level/ Subproject | ABCD-reports Burndown chart | Progress meeting with customer Risk register Subproject meetings Top Five Risks | Ad-hoc handling of risks Code Review Estimation Integrations Project plan Early testing Technical debt Tasks to other subprojects | Task prioritization |
| Work package level | Issue board Risk matrix | None identified | | Problem analysis using A3s No Gold plating Pair design Retrospective |

As the interviews were conducted in Norwegian, the analysis was performed by the first author of this article with help from one researcher (see acknowledgement). The second author contributed in writing the theory section and discussing findings and contrasting this with findings from prior literature. The discussion was developed in a series of phone meetings. Results from the analysis were presented to some of the informants for validation.

## 4.3 Results

### 4.3.1 Case description

The project developed a product which consists of hardware and software components and the development has taken more than ten years and for a long period involved about 200 engineers. Main subprojects were for hardware, for software, and for integration and testing. The project had more than 50 subcontractors. The product had strict non-functional requirements for performance, security and safety, and a set of functional requirements which were mainly known in early phases of development. Most requirements were defined early in the project and a contract with the client was written based on these requirements. The project was managed as a traditional project using a V-model but followed agile development methods primarily for the software components (a first-generation large-scale agile project). The project was divided into six subprojects, with up to six work packages. The work packages had a work package leader and a team. Some of the software teams used the Scrum development process with two-week iterations, starting with iteration planning and ending with an iteration review. There were numerous dependencies between the subprojects and these dependencies were both technical and organisational.

### 4.3.2 Project risks

When asked to assess risks after the framework suggested by Ropponen and Lyytinen (2000), the informants rated "scheduling and timing risks" and "system functionality risks" as the two most important. One respondent stated, "the last two years, we have been pressed hard on time and cost . . . it has been a large focus on just implementing what must be implemented" (work package team member). Another respondent was asked what was most important and stated that it was "functionality and performance . . . but it is always connected to schedule".

Risks are mainly interpreted as threats, but sometimes also opportunities.

*We have tried to talk about threats and opportunities and uncertainty, but we notice that we easily fall back on talking about threats. (Manager)*

However, some state that the reporting structure of the project focuses on reporting benefits from work package level and up (see more on ABCD-reporting later), and people report things like:

*this is an opportunity – here we have a chance to work more efficiently. (Work package manager)*

### 4.3.3 Explicit and implicit practices at different levels of the project

We show the mitigation practices to manage risks at the project/subproject level and show the practices we characterise as explicit and implicit in Table 4.2. The table also shows which practices were common to both project/subproject and work package levels, and which were specific. In the following sections, we provide descriptions of each of these practices.

### 4.3.4 Explicit uncertainty mitigation

#### 4.3.4.1 Common

*ABCD-reports:* Every second week the project managers wrote a report, called the "ABCD-report", indicating what had been **A**chieved, resulting **B**enefits, **C**oncerns and what they planned to **D**o next. The subproject managers received the ABCD reports from their work package leaders, and risks were reported as concerns. It was possible to indicate at what level a risk should be handled. Another option was to present issues requiring concrete actions directly to the project manager in project meetings.

*Burndown charts:* These were used at subproject and work package levels. At work package level, there were multiple variants of burndown charts – as a board, in different software tools which changed during the course of the project.

*Issue boards:* Issue boards were used to register and follow up risks, bugs, problems and improvement suggestions at subproject and work package levels. In a work package, one informant stated:

> this is something we use for a number of purposes [. . .] we meet once a week and discuss what has been registered. (Work package manager)

*Risk matrix:* The risk matrix was established early at project and subproject levels. The matrix was updated before progress meetings described below, but also on a need-be basis if something came up. Although it was not mandated by the project managers, some work package leaders discussed risks at work package level and would bring them into the risk matrix at subproject level. An informant stated that:

> work package leaders are clearly involved in developing the risk matrix at subproject level. (Work package team member)

### 4.3.4.2 Specific for project/subproject level

*Progress meeting:* Three to four times a year, the project had a meeting on progress with the customer. They walked through the risk matrix and risk register as preparation for the meeting. This was done on project level, based on preparations in all subprojects where subproject managers assessed their risk registers.

*Risk register:* The program established a risk register in the beginning of the program by brainstorm meetings at subproject level. The items in the register were given a probability and a consequence. A person was given responsibility of mitigating actions. Status of actions were discussed with the customer at the progress meetings.

*Subproject meetings:* Risks were discussed in subproject meetings, where the subproject managers met with work package managers. This was based on discussions between work package team leaders and team members, not necessarily in a separate meeting on this topic.

*Top Five Risks:* Every subproject manager brought a list of "top five" internal risks to a project meeting, and the management discussed the risks and developed a list of the project "top five". This list was communicated internally through the project Intranet. In our interviews, this was only mentioned by people in management positions, the awareness of this list might be low in the project organisation.

### 4.3.4.3 Specific to work packages

From our interview material, we could not identify explicit practices that were specific to work packages. There were no obligations for explicit practices at this level.

## 4.3.5 Implicit uncertainty mitigation

### 4.3.5.1 Common

*Ad-hoc handling of risks:* Several informants said risks were handled ad-hoc as they were discovered.

> *What can we do when things stop? . . . we find solutions to ensure that we deliver as planned (Manager).*

Another informant stated that at work package level,

> *if something shows up, we call for a meeting right away, but there is not necessarily any process or description of it. (Work package manager)*

*Code Review:* Many said they did multiple code reviews both in the work packages and in subprojects. Sometimes code was reviewed by several peers if it was a large or complicated part. If it was a minor change then just one person might review it. One work package had a practice of having at least two people reviewing the code, as code quality was particularly critical in this part of the project.

*Estimation:* Because of the complexity and innovations of the tasks, accurate estimations were one of the major challenges throughout the project. The work tended to be underestimated, in particular the work regarding integration of hardware and software components. At work package level, tasks were individually estimated in man-hours. In one work package, they tried group estimates, but it was not a practice that they maintained.

*Integrations:* The project integrated deliverables from all subprojects at certain intervals, but because some deliverables include hardware, the increments could be very large, in the order of one to two years of work (while many of the software work packages used two-week iterations).

> *This means that groups that consist of system engineers, software architects and software developers work on the same functions, but not at the same time. . . . when we pick up a topic, the system engineers have forgotten what they talked and thought about. (Software developer)*

Although many expressed that integrations were too infrequent, they described integrations as a major factor to reduce risks:

> *because you go through the whole process over a short period and can find uncertainties in the whole V. (Work package team member)*

Some said that it had been difficult to integrate at the planned milestone dates, because of the difficulty to synchronize parts from so many people at an agreed milestone:

> *We are in a situation now, where we are depending on a delivery from another work package, they do not know exactly when they can deliver the first version, and we know there will be errors. So, we need to test, which will take time, and we are unsure about how long. Then someone else is waiting at the other end to make use of the delivery. (Work package manager)*

*Project plan:* Some risks were identified during planning, for example when the project team needed to commit to dates and milestones.

*Early Testing:* Taking up agile practices such as early integration led to much earlier testing. In previous projects testing was only done at the end of the project. The early testing led to a reduction of risks.

*Technical debt:* Previous studies on risk management in agile projects identified technical debt as one important concern. In our case, the informants expressed a varying degree of awareness on technical debt. One said that he was not aware of the term, but has experienced that

> *we cannot always perform tasks the way we want to, there is not enough time for that. (Work package team member)*

Other informants reported that they had frequent discussions on technical debt and registered all debt. The debt was then regularly assessed. The technical debt had so far not been shown in the risk matrix, they talk about technical debt, but had not identified it as risk.

*Tasks to other subprojects:* Tasks were regularly assigned from one work package to another and they tried to make dependencies visible. In particular this was important when limited resources could be assigned to certain tasks, especially if they were considered high priority tasks.

### 4.3.5.2 Specific: Project/subproject level

*Task prioritization:* Some stated that the project was in a better position to manage risks compared to previous projects due to the uptake of agile methods and frequent re-prioritization and focus on work tasks. According to the project manager: "The method makes you have more focus". A subproject manager explained that they had influence on prioritisation in other subprojects:

> *we have worked with getting them to understand what is important to us. (Manager)*

#### 4.3.5.3 Specific: Work package level

*Problem analysis using A3s:* A technique borrowed from lean manufacturing used to document a problem on an A3 sheet, including: Background, what has been done regarding the problem and a conclusion. This practice was used by the work package teams in this phase of the project. These were stored to give background if a problem reoccurred.

*No gold plating:* Most informants expressed that over-investing in quality or functionality was not a problem in the project.

> The last two years, we have been pressed hard on time and costs, so towards the end of the project there has been a strict focus on just implementing what needs to be implemented. (Work package team member)

The work package leader would be informed of the team's activities and ensured that activities were within scope. According to a work package member:

> we focus on being finished rather than the "nice to have"-stuff. (Work package team member)

*Pair design:* Both pair design and pair programming have been used to some extent, but there was no project-wide policy on the use of these techniques. Pair programming allows code review to be done while writing the code, which is believed to increase the code quality. Also, discussions while doing development could lead to more optimal design decisions which could reduce risks related to code quality.

*Retrospective:* Some of the work packages conducted regular retrospectives, while others did not. Some that did retrospectives combined this with a planning meeting where most of the time was spent thinking ahead, and little on discussing lessons learned during the last iteration. Retrospectives could be used to discuss estimation precision, but our informants did not express such use of these meetings.

## 4.4 Discussion

We return to our research question, *how does a large software/hardware development project using agile practices manage uncertainty at project/subproject and work package levels?*

The case studied was a large project developing a product consisting of hardware and software components. The software engineering field has experienced many changes in recent years due to agile development methods which bring in new practices and new terminology such as «technical debt». The project has taken up agile methods through more frequent integration of components than in previous projects as well as in using agile development methods mainly at work package level. We described the case project as a first-generation large-scale agile development project.

The results show a total of 21 uncertainty mitigation practices, where we have classified eight as explicit and 13 as implicit. Further, five practices are specific to the project and subproject levels (four explicit and one implicit), while four are specific to the work package level (all implicit). This finding echoes findings on coordination in large-scale agile development, where studies show that there are far more practices in use in successful projects than are described in recommendations in existing frameworks (Dingsøyr et al., 2018b).

Although there were a large number of practices, their use varied much between work packages and between subprojects. For example, estimation techniques are used to varying degrees, the awareness of «technical debt» was strong in some work packages while others were unaware of the term, and the agile practice of retrospectives was used to varying degrees. Some of the explicit practices such as the risk matrix and risk register were used across the subprojects and also at some work packages. Our study is to our knowledge the first to identify practices at different organisational layers in projects. We also identified a larger variety of practices than can be seen in empirical studies of small agile development projects (Dönmez & Grote, 2018; Elbanna & Sarker, 2016; Siddique & Hussein, 2016) or from studies of uncertainty management in general.

An interesting observation was that at work package level there were no specific explicit practices. All explicit practices had been deployed at project level and most of these techniques required some flow up or down between the hierarchy of the project. This was different for the implicit practices where multiple specific techniques were used at the work package level. The specific implicit practices at work package level were primarily used to manage "system functionality" risks (Ropponen & Lyytinen, 2000) and consisted of practices from agile development such as retrospectives and pair programming and from lean production with the use of A3s to document problems and solutions. In contrast, the risk management techniques used at project level were focused mainly on "scheduling and timing risks" (Ropponen & Lyytinen, 2000).

The interviewees did not make distinctions between risk and uncertainty management. However, going back to the definitions presented in the literature review section, it became clear during the data analysis that the project was focusing mainly on risks (i.e., possible events which might impact the project) using traditional explicit risk management techniques deployed throughout the project. As in previous findings, this program focused their risk management primarily on threats and not on opportunities. In comparison, the work packages were focusing primarily on implicit techniques for uncertainty management (i.e., their inability to determine the scope and the task durations precisely for the whole project).

Previous studies of risk management in agile development have identified challenges with separation of development and operations and a growing technical debt (Elbanna & Sarker, 2016). The separation of development and operations was not directly addressed in the mitigation practices identified in our study, while some brought this topic up in interviews. Technical debt was a topic with varying practices across work packages. For large projects, a crucial topic identified in previous studies has been coordination between different teams. This was mainly done through a traditional project organisation in our case, and for example not through meetings such as Scrum of Scrums. However, the informants stated that "scheduling and timing" were the main risks in the project, and this was much related to coordination between teams. A practice to manage this type of uncertainty was to give "tasks to other subprojects".

## 4.5 Conclusion

This study shows how uncertainty is managed in a large project with several subprojects, using a combination of practices from project management and from agile development methods. In line with previous findings, we found that uncertainty management practices were mainly focusing on handling threats, and to a smaller degree on opportunities.

We are not aware of previous studies describing how uncertainty is managed at the different levels in large projects. The main contribution of this study is that our case shows the high number of practices in use, some are used on all levels, while others are used only the project/subproject level or at the work package level.

The use of many of the practices vary, for example some work packages were very conscious on registering technical debt, while others did not use this term. The specific practices at project/subproject level are mainly practices that explicitly handle uncertainty while the practices at work package level are practices that implicitly handle uncertainty. Further, some of the practices link between the layers in the project, such as the ABCD reports, the issue boards and the registration of technical debt.

With the increasing importance of software in many large projects, we believe more projects will have to manage layers of risks with a wide range of practices in the future. This exploratory study highlights characteristics of uncertainty mitigation in first generation large-scale agile projects that blend "traditional" project management and agile methods.

## 4.6 Limitations

The major limitation of this study is that it is an exploratory study with a limited data collection. The case was selected as a "typical" case of a first-generation large-scale agile development project. An "extreme" case with more effort needed on uncertainty management might have opened up other topics for further investigation.

The interviews were done during the third of four project phases, rather than longitudinally over the life of the project. Given the type of study, with semi-structured interviews of informants on how uncertainty was managed, we do not have detailed information on use of all practices identified.

## 4.7 Implications for theory and practice

We have presented findings from a case study of a first-generation large-scale agile development project and what practices were in use for uncertainty management at two levels. This study gives important directions when attention is given to second generation frameworks which are increasingly used in the software industry. In particular, we would like to highlight five directions:

First, a number of challenges have been identified in large-scale agile development (see for example (Bass, 2019)) which includes topics such as interteam coordination, technical architecture and assigning priority to user needs. How do the resolution strategies suggested in the various second-generation large-scale agile methods work in practice?

Second, an implicit practice to manage uncertainty in agile development is to assign work to teams. A team is more robust than individuals but studies of large-scale development suggests that teamwork in this context is different from teamwork in single-team settings (Lindsjørn et al., 2018). How can second-generation frameworks facilitate good teamwork while managing project uncertainty?

Third, the technical architecture is a major source of uncertainty. Prior studies of first generation large-scale agile development suggests that the architecture has been stable (Dingsøyr et al., 2018a) while advice is to work more iteratively also with architecture (Nord et al., 2014). How do projects address this type of uncertainty when using second generation methods?

Fourth, large-scale agile development projects are organised as multi-team projects, often with a structure as in our case with subprojects and work package teams. What alternative linkage strategies exist, and how are these strategies used in practice?

Finally, we think there is a need to revise existing published risk frameworks such as (Ropponen & Lyytinen, 2000) in order to fit needs in large projects both for first- and second-generation large-scale agile development methods.

Although this is an exploratory study primarily intended to generate directions for new research, we believe the description of uncertainty management practices in the case has several implications for practice:

First, the case study shows a large number of explicit and implicit practices in use to manage uncertainty. Prior studies on coordination in large-scale agile development projects have also found a large number of arenas for coordination. It seems reasonable that large projects will need a large number of practices to manage the diverse range of uncertainty during project execution.

Second, we believe the list of practices identified in Table 4.2 can be valuable when planning new large projects or seeking inspiration to manage certain types of uncertainty during project execution.

**Acknowledgements:** This article was written with partial support from the Agile 2.0 project, supported by the Research Council of Norway through grant 236759 and companies Equinor, DNV GL, Kantega, Kongsberg Defence & Aerospace, Sopra Steria and Sticos. We are very grateful to Andreas Bredesen and Elin Lintvedt at Kongsberg Defence & Aerospace for administering a project-wide survey and setting up individual interviews. Further, we would like to thank Finn Olav Bjørnson at the Department of Computer and Information Science at the Norwegian University of Science and Technology for helping with the analysis of the interview material. A prior version of this paper was presented at the IRNOP2018 conference, we are grateful to comments from conference and book reviewers and to conference participants for discussions. We would also like to thank Agnar Johansen at the Norwegian University of Science and Technology for discussions on study design and input on the interview guide. We are further grateful to Amany Elbanna at the University of London for comments on the interview guide and discussion of findings.

# Appendix 1: Interview guide

Will use "uncertainty management" as managing *"uncertain event or conditions that, if they occur, has a positive or negative effect on one or more project objectives such as scope schedule, cost and quality" (PMI Body of Knowledge).*

## Part 1

**Background**

1. Could you describe your role in the project?
2. Could you describe the work method in your part of the project?
3. Are your tasks affected by other work packages or subprojects? How?
4. Do you work in a distributed team? Hardware or software?

**Uncertainty management**

[Risks]
5. Do you identify risks to project progress in your part of the project? What hinders project progress in your part? Are there particular challenges you are facing?
6. How do you work to identify risks? (practices? tools?) what do you normally do about them (issues/ challenges/ problems)?
7. Are you aware of (implicit) practices that reduce risks to project progress? Is there something done on the work package/subproject/project level about them?
8. How are risks communicated in the project organization? Do you think your technical lead or PM is aware of this? Are they on denial or can/do something about it?
9. How do you work to identify opportunities? (practices? tools?)

[Questions about identified risks in agile projects]
10. Would you say that the product has technical debt? How much debt you think the team accumulates? Do you think there is a way to pay part of this debt back during the development? Do you think some could be paid back after the project? How?
11. Do you have established approached to manage technical debt? Do you keep a log or any other way to keep track of these debts?
12. Are there challenges in handing over products from development to operations/maintenance?
13. What PM tools are you using? Is this available for all teams in the organisation? How are project management tools used in the project? Alignment?
14. Is knowledge regarding work tasks preserved for later maintenance? Have you been involved in system upgrade or maintenance tasks?

[Opportunities]
15. Are you aware of (implicit) practices to identify opportunities in the project?
16. How is this handled?
17. How are opportunities communicated in the project organization?

18. Do you estimate the effort involved in work tasks? How? (practices? tools?)
19. How is progress on work tasks communicated to other parts of the project?
20. Do you see potential for improvement for managing
    * Risks?
    * Opportunities?

## Part 2

Questionnaire: Project risks

Please indicate how important the following risks are in your current project. Use the following scale:
1 – Not important
2 – Slightly important
3 – Moderately important
4 – Important
5 – Very important

Please indicate how important you think the following risks will be in future similar projects. Use the following scale:
1 – Not important
2 – Slightly important
3 – Moderately important
4 – Important
5 – Very important

*Risks to be assessed (from Ropponen and Lyytinen (2000)):*

Scheduling and timing risks
System functionality risks
Subcontracting
Requirement management risks
Resource usage and performance risks
Personnel management risks